\documentclass[prl,amsmath,twocolumn,superscriptaddress,a4paper]{revtex4-1}
\usepackage{graphicx}
\usepackage{color}
\usepackage{amssymb,braket}
\usepackage{ulem}

\definecolor{green}{rgb}{0,.5,0}

\begin{document}

\title{Cascading Proximity Effects in Rotating Magnetizations}
\author{Thomas E.~Baker}
\affiliation{Department of Physics \& Astronomy, California State University Long Beach, Long Beach, CA 90840, USA}
\affiliation{Department of Physics \& Astronomy, University of California, Irvine, Irvine, CA 92697, USA}
\date{\today}
\author{Adam Richie-Halford}
\affiliation{Department of Physics \& Astronomy, California State University Long Beach, Long Beach, CA 90840, USA}
\author{Ovidiu E.~Icreverzi}
\affiliation{Department of Physics \& Astronomy, California State University Long Beach, Long Beach, CA 90840, USA}
\author{Andreas Bill}
\email[Correspondence should be directed to: ]{abill@csulb.edu}
\affiliation{Department of Physics \& Astronomy, California State University Long Beach, Long Beach, CA 90840, USA}
\affiliation{Instituto de Ciencia de Materiales de Madrid, CSIC, Cantoblanco, 28015 Madrid, Spain}
\begin{abstract}
We demonstrate two effects that occur in all diffusive superconducting-magnetic heterostructures with rotating magnetization: the reappearance of {\it singlet} $\ket{0,0}$ correlations deep in the magnetic material and a cascade of $s=1$, $m=0,\pm1$ components (in the two spin$-1/2$ basis $\ket{s,m}$). We do so by examining the order parameter and Josephson current through a multilayer with five mutually perpendicular ferromagnets. The properties of the middle layer determine whether the current is due to $m=0$ or $m\neq 0$ contributions. We conclude that so-called long- {\it and} short-range components are present across a proximity system with rotating magnetization.
\end{abstract}


\maketitle

The order parameter of Cooper pairs tunneling from a singlet pairing superconductor (S) into a magnetic material has a distinctive character whether the magnetization is homogeneous or inhomogeneous.  In a homogeneous ferromagnet (F), pairs tend to break apart due to the exchange field and may acquire a net momentum to become a triplet as a result of the Fulde-Ferrell-Larkin-Ovchinnikov (FFLO) effect \cite{larkinJETP64,fuldePRL64}. In this case, the spin states of the Cooper pair have zero projection along the quantization axis, $m = 0$ \cite{buzdinRMP05}. If the magnetic material is inhomogeneous, the presence of rotating magnetization (domain walls) will further introduce components with $m \neq 0$ as first pointed out in Refs.~\cite{bergeretPRL01,bergeretRMP05}.

One important difference between the various components is the spatial decay range of their exponential envelope. In a diffusive system, the decay of the $m = 0$ components is determined by the ferromagnetic coherence length, $\xi_F = \sqrt{D_F / h}$, ($D_F$ is the diffusion length and $h$ the exchange field of the ferromagnet) \cite{buzdinJETP82,buzdinRMP05}, while the $m \neq 0$ components decay with the characteristic normal coherence length $\xi_N = \sqrt{D_F/2\pi T}$. Under typical experimental conditions, $h\gg T_c \geq T$, where $T_c$ is the superconducting critical temperature of the proximity system. Defining $\xi_c \equiv \sqrt{D_F/2\pi T_c}$, the condition $\xi_F\ll\xi_c\ll \xi_N$ is thus generally satisfied. For this reason, $m=0$ and $m \neq 0$ are often termed short- and long-range components, respectively. It is commonly assumed that the short-range components exist only very near interfaces while the long-range triplet components penetrate deep into the magnetic material.

In this Letter we demonstrate, using numerical solutions of the Usadel equations \cite{usadelPRL70} for multilayers and the derived, exact analytic solution of the wide, homogeneous F, that in diffusive inhomogeneous magnetic systems the distinction between short- and long-range components is inadequate since {\it both} $m=0$ and $m\neq 0$ components are always present across the material and may constitute a non-negligible amount of experimentally measured current. Importantly, this effect occurs in inhomogeneous magnetic structures even when the magnetization rotates over long enough distances to exclude antiferromagnetism as the reason for the survival of singlet components \cite{bergeretPRL01b}. We propose an experiment where the existence of singlet and triplet components can be independently probed deep in the magnetic material.

\begin{figure}[h]
\includegraphics[width=\columnwidth]{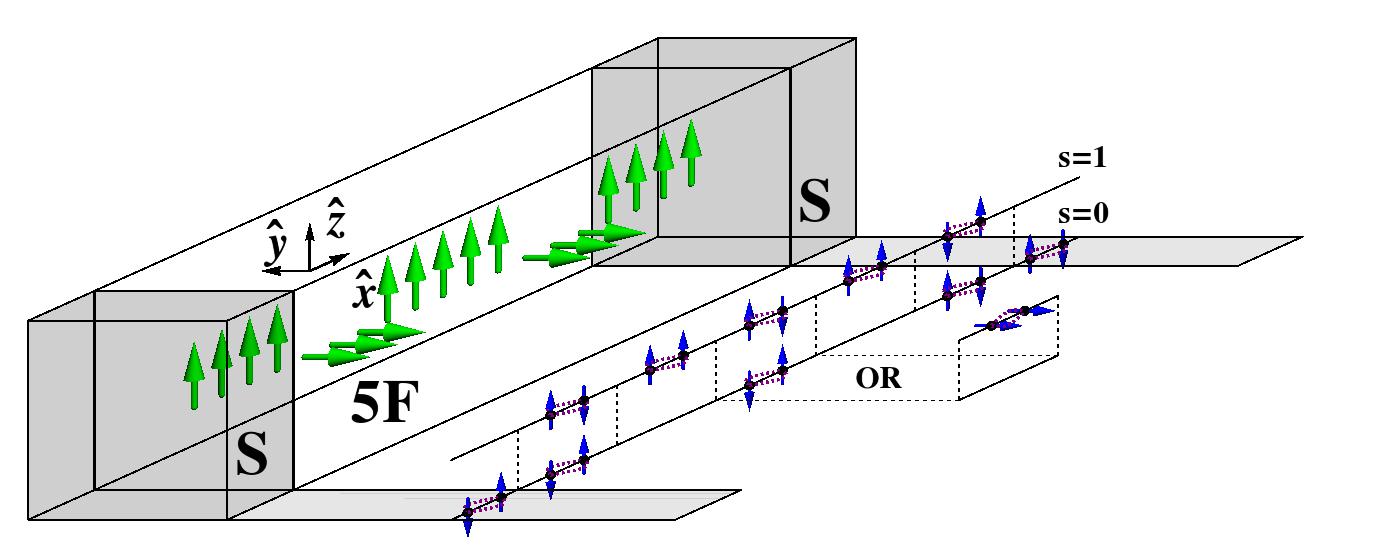}
\caption{\label{Figure1} (color online) Structure and magnetization of the S5FS multilayer and schematic representation of the dominant components of the order parameter (linked spin arrows) in the various layers. Singlets ($m=0$) re-emerge in the central layer (even for $d_F\gg \xi_F$).}
\end{figure}

The mixing of $m=0$ and $m\neq 0$ components across a material with rotating magnetization is best revealed in a multilayer of $\nu$ misaligned homogeneous Fs embedded in a Josephson junction. A cascade effect is observed where the components are transformed into another combination and back again as one crosses interfaces between misaligned Fs. We consider in particular $\nu = 5$ perpendicularly oriented magnetic films  as schematically depicted in Fig.~\ref{Figure1}. The interest of the S5FS system lies in the possibility of a $\mathbf{\hat{y}}$ component appearing in the middle layer and the prediction that in some limits, the measured Josephson current has the characteristic signature of an $m=0$ component while other configurations will see a current expected from the $m\neq 0$ nature of the order parameter.

To illuminate the results of the numerical calculation, we also present a complete analytic solution of Usadel's equation for a homogeneous F to observe that the same mechanism that converts singlet Cooper pairs into $m= 0$ triplet components (FFLO effect) also converts this component back into singlet pairs through a ``reverse FFLO effect" if an adequate exchange field is applied.
\begin{figure}[t]
\includegraphics[width=\columnwidth]{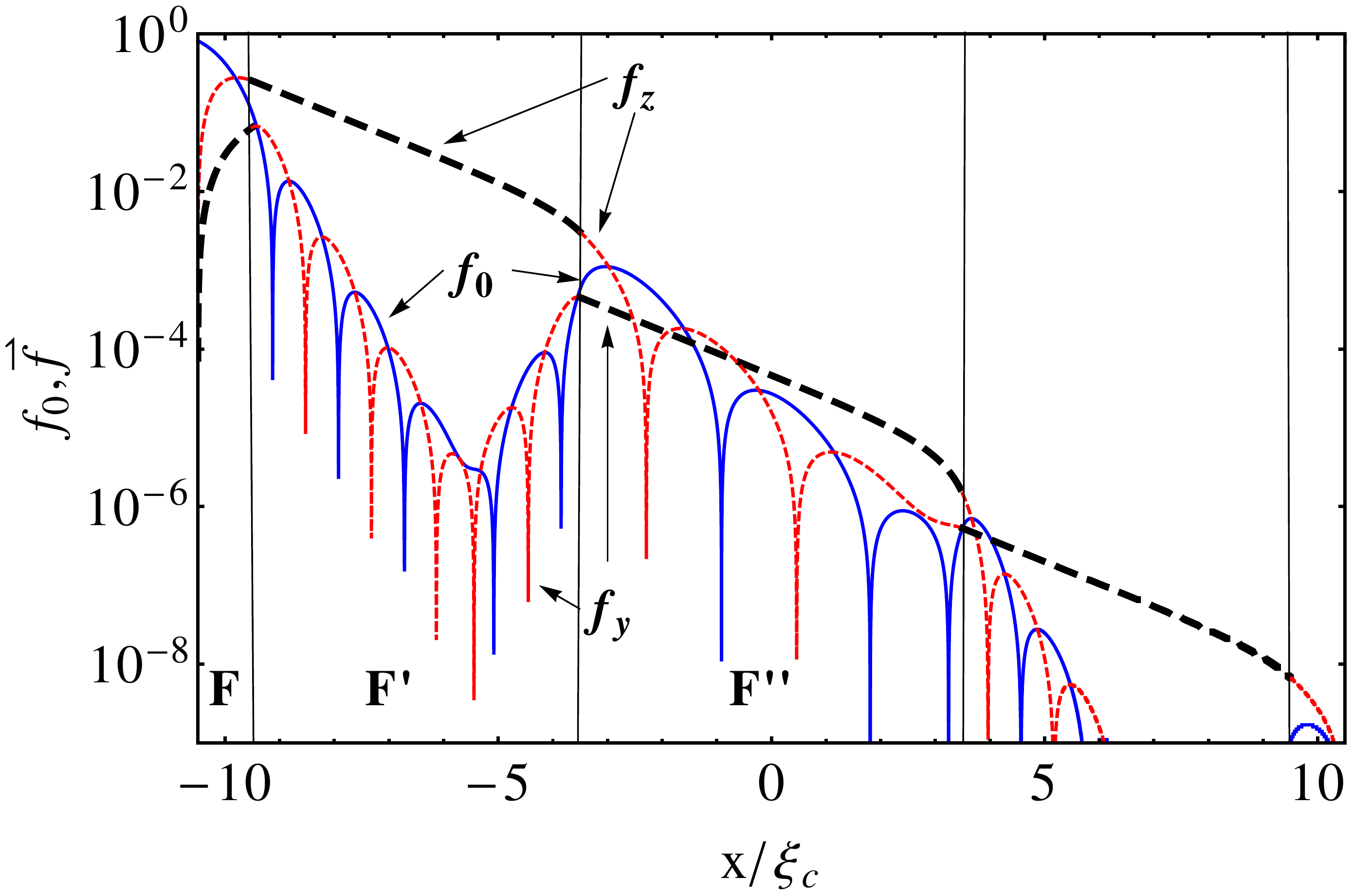}
\caption{\label{Figure2}
(color online) Order parameter for the S5FS multilayer depicted in Fig.~\ref{Figure1}. The Gor'kov functions found numerically in the five Fs (separated by light vertical lines) are generated by the proximity of the S located on the left of the figure frame. Noteworthy is the reappearance of sizable {\it singlet} contributions deep in the multilayer due to the cascading effect of $m=0$ and $m\neq 0$ components. In all layers, the solid (blue) and dotted (red) lines represent the $m=0$ singlet and triplet components respectively, while the dashed (black) line is the $m \neq 0$ component.
Exchange fields are set to $h_i=(3,14,3,14,3) \pi T_c$ in the 5F system and thicknesses of the layers are $d_i=(1,6,7,6,1)\xi_c$. $\omega_n = \omega_0$ and $T = 0.4 T_c$.}
\end{figure}

We conduct our analysis in the diffusive limit where the system is described by Usadel's equations for the Green functions $g_0, \mathbf{g}=(g_x,g_y,g_z)$ and the Gor'kov functions $f_0,\mathbf{f}=(f_x,f_y,f_z)$ including all possible spin correlations of pairs in the F. The scalar $f_0$ describes the singlet, and the vector $\mathbf{f}$ describes the triplet components. 
In the Ivanov-Fominov trigonometric parameterization ($f_0=M_0\sin\vartheta$ and $\mathbf{f}=-i\mathbf{M}\cos\vartheta$) and the Matsubara formalism ($\omega_n=\pi T(2n+1)$, $n\in\mathbb{Z}$)  Usadel's equations in F read ($M_0^2-|\mathbf{M}|^2=1$) \cite{ivanovPRB06,ivanovPRB09}
\begin{eqnarray}
\label{usadeltheta}
\frac{D_F}{2}\boldsymbol{\nabla}^2\vartheta - M_0 \omega_n \sin\vartheta - (\mathbf{h} \cdot \mathbf{M}) \cos\vartheta &=& 0,\\
\label{usadelM}
\frac{D_F}{2}\left( \mathbf{M} \boldsymbol{\nabla}^2M_0 - M_0\boldsymbol{\nabla}^2\mathbf{M}\right)
+  \mathbf{M} \omega_n \cos\vartheta &&\\
- \mathbf{h} \,M_0 \sin\vartheta &=& 0.\nonumber
\end{eqnarray}

We consider the wide limit which allows the solutions for the S$\nu$F and $\nu$FS proximity systems to be summed to determine the order parameter in the S$\nu$FS Josephson junction \cite{zaikinZP81}.  We further assume that the pairing potential $\Delta$ is constant up to the SF interface which neglects the inverse proximity effect, changes in the critical temperature, and certain effects relevant inside one coherence length \cite{vasenkoPRB08,golubovRMP04,richardPRL13}. The Usadel equations are solved numerically with transparent interfaces and boundary conditions $(M_0, \mathbf{M}) = (1,\mathbf{0})$, and $\vartheta = \arctan(\Delta/\omega_n)$ at the SF (FS) interface, while the functions are assumed to vanish at the right (left) end of the S$\nu$F ($\nu$FS) multilayer.

The Gor'kov functions in the F multilayer of Fig.~\ref{Figure1} are shown in Fig.~\ref{Figure2}. The most noteworthy feature of this calculation is the presence of sizable $m=0$ components in all five layers even though the total width of the magnetic system is about $20\xi_c$. To ease the reading of the figure, we assign the solid blue (dotted red) line to the $m = 0$ singlet (triplet) components and the long dashed black line to the $m = \pm 1$ triplet components.

To understand the physical content of Fig.~\ref{Figure2} and the cascading effect, let us focus first on the F'F" interface. The dashed (black) line in F' is the "long-range triplet component" predicted in Ref.~\cite{bergeretPRL01}. Entering the F" layer, which has a magnetization along the $\hat{\mathbf{z}}$ direction, this component transforms back into an $m=0$ component ($f_z$ is along a direction parallel to the magnetization in F").  This process is the reverse of the generation of the $m\neq 0$ component at the FF' interface. The interesting aspect is that the process is necessarily accompanied by the generation of a {\it singlet} component through a reverse FFLO effect. This component is depicted as the solid blue line in F". A similar effect is observed throughout the multilayer resulting in a cascade of $m=0$ and $m\neq 0$ components generated by each rotation of the magnetization.

To better understand this cascading effect and the generation of singlet components deep in the magnetic multilayer, we digress from the description of Fig.~\ref{Figure2} to discuss an analytic solvable model where a similar phenomenon appears. We return to the discussion of Fig.~\ref{Figure2} below.

Consider Eqs.~(\ref{usadeltheta},\ref{usadelM}) in the case of a single, homogeneous magnetic film and assume that only $m = 0$ components are present. The equations take the well-known form \cite{zaikinZP81,houzetPRB05,vasenkoPRB08,faurePRB06}
\begin{equation}\label{usadeleq}
D_F \partial_x^2\theta = 2\left( \beta + \cos\theta/\tau_s \right) \sin\theta
\end{equation}
with $\beta\equiv \omega_n+i\mathrm{sgn}(\omega_n)h$ and $\theta = \theta_r + i \theta_i$. We include the spin flip scattering time, $\tau_s$, to take into account the possible presence of magnetic impurities \cite{houzetPRB05,ivanovPRB06,crouzyPRB07}.

It is well established that for a homogeneous F with magnetization along $\mathbf{h}= h {\bf\hat z}$, adjacent to a singlet superconductor, the Gor'kov function reads $\mathcal{F} = f_0 + i f_z$ and both components are real functions of position \cite{ivanovPRB06}.  Introducing the parametrization $\mathcal{F} = \sin\theta$ with $\theta$ complex, the imaginary (real) part of $\mathcal{F}$ can be identified with the $m = 0$ triplet (singlet) components $f_z$ ($f_0$).  This parametrization is connected to that of Ivanov-Fominov  by defining $\theta_r = \vartheta$, $M_0 \equiv \cosh \theta_i$ and $M_z \equiv -\sinh \theta_i$. 

The exact solution of Eq.~\eqref{usadeleq} can be found by using the mathematical equivalence of the equation with that of the classical mechanical system known as the bead on a hoop.  The complete, closed form solution for the latter system has been derived by two of the authors in Ref.~\cite{bakerAJP12}. The solution for the equation without spin-flip scattering has been presented in Ref.~\cite{bakerJSNM12} and shown to agree with approximate and numerical results \cite{houzetPRB05,vasenkoPRB08,ryazanovPRL01,ARH}.
The use of the closed, analytic form of Ref.~\cite{bakerAJP12} for the present problem requires the generalization to the complex quantity $\beta$ \cite{linksolAJP}. The exact solution of Eq.~\eqref{usadeleq} is 
\begin{equation}\label{usadelsol}
\theta(x)=2\arctan[\Omega\;\mathrm{sech}(x\;\Omega\sqrt{-2\beta/D_F}-\Gamma)]
\end{equation}
where $\Omega=\sqrt{-1/\tau_s\beta-1}$ and $\Gamma=\mathrm{arcsech}[\tan(\theta_B/2)/\Omega]$.
\begin{figure}[h t]
\begin{center}
\includegraphics[width=5cm]{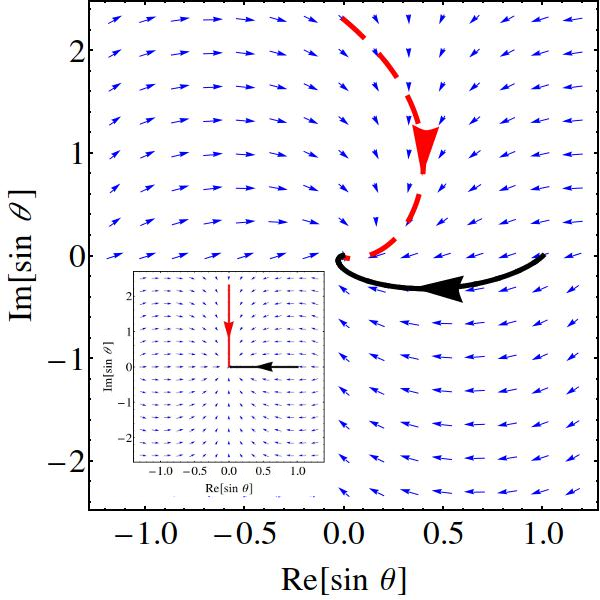}
\end{center}
\caption{\label{Figure3} (color online) Real and imaginary parts of the Gor'kov function $\mathcal{F}=\sin\theta$, Eq.~\eqref{usadelsol} with $\omega_n=\omega_0$, and its gradient ${\mathbf\nabla}\mathcal{F}=(\partial\mathcal{F}/\partial\theta_r,\partial\mathcal{F}/\partial{\theta_i})$ at each point (normalized small arrows) represented on the Argand plane for a homogeneous film. The thick lines represent the order parameter in the F layer when starting with a pure real (solid black) or imaginary (dashed red) value, corresponding respectively to the singlet and triplet states at one boundary of F. The curves are parametrized by the position $x$ in the F and spiral towards the origin since the order parameter decays in the F. ${\cal F}$ {\it always} develops into a singlet-triplet mixture in F (see text). $h=3\pi T_c$. Inset: The same in a normal metal N ($h=0$).}
\end{figure}

The general behavior of Eq.~\eqref{usadelsol} is shown in Fig.~\ref{Figure3} where we plot $\mathcal{F} = \sin\theta$ on the Argand plane. The small arrows show the flow of the order parameter as one moves into the F.  The two paths (thick solid black and dashed red lines) correspond to the FFLO and reverse FFLO states.  They are parametrized by the position $x$ in the F.  The first case, depicted as the solid line that begins on the real axis, was obtained by placing a real value of the order parameter into Eq.~\eqref{usadelsol} at the left boundary of F ($x=0$).  The solution immediately generates an imaginary contribution to $\mathcal{F}(x)$ as one moves into the F. This is the signature of the $m = 0$ triplet component due to the FFLO effect.  The other (dashed) curve is given a pure imaginary order parameter at the left boundary and also immediately acquires a real part and thus a singlet component in the F (reverse FFLO effect).  The small arrows show that, irrespective of the composition of the order parameter at the left boundary of the homogeneous F (the starting point of the thick curves), the flow of the solution always leads to a complex order parameter in F. This contrasts with the behavior of the order parameter in a non-magnetic metal presented in the inset of Fig.~\ref{Figure3}: a pure singlet (triplet) component at the left boundary remains singlet (triplet) across the material. 

The reverse FFLO effect represented by the dashed thick curve of Fig.~\ref{Figure3} is particularly relevant for Fig.~\ref{Figure2} since it shows that the re-emergence of the singlet component in the F layer is not a product of the geometry we consider but an inherent property of Eq.~\eqref{usadelsol} or Eqs.~(\ref{usadeltheta},\ref{usadelM}). The singlet component in F" of Fig.~\ref{Figure2} necessarily accompanies the transformation of the $m\neq 0$ component in F' into the $m=0$ one in F". The analytic model and Fig.~\ref{Figure3} also emphasize the importance of the boundary conditions for the composition of the order parameter inside the F.

Having motivated the appearance of the singlet component (solid blue line) in F" in Fig.~\ref{Figure2} with the analytic model, we proceed to explain the  cascading effect. We find it useful to consider the {\it Gedankenexperiment} where each layer is examined separately with {\it effective} boundary conditions. Numerically, we calculated the order parameter simultaneously at all points of the multilayer, setting boundary conditions at the SF interface on the left and at the outer right edge of the fifth magnetic layer of S5F. However, for the purpose of describing the cascading effect and using the insight brought by the analytic model, we consider the influence of the Gor'kov functions $(f_0,\mathbf{f})$ at the boundaries of each F to discuss the order parameter in the F. Since the differential equations are second order the values of the functions at both boundaries of a given F determine a unique solution inside that F.

Using this approach, the $m \neq 0$ component $f_z$ (dashed black line) in F' is for example readily understood from the fact that the $m = 0$ triplet component (dotted red line) in F contributes to the left effective boundary condition that generates the $m\neq 0$ component in F' \cite{bergeretPRL01}. At the F'F" interface the magnetization is rotated again. The order parameter at that interface is a different mixture of Gor'kov components and thus forms another left effective boundary condition for F". As motivated by Fig.~\ref{Figure3} the effective boundary condition at the F'F" interface generates $m = 0$ components in F". Similar effects are then observed in the next layers.

Consider now the effective boundary condition on the right of each F layer. That boundary condition explains for example the existence of an extra $m\neq 0$ component (dashed black line) in F. By the same mechanism as described above the $m=0$ singlet and triplet components of the order parameter at the FF' interface generate the $\mathbf{\hat{y}}$, $m\neq 0$ state in F. Given the actual boundary condition at the SF interface the latter component must necessarily vanish at that interface.

We conclude from the above reasoning that a cascade effect is observed by which the rotation of the magnetization at each interface generates a new linear combination of $m=0$ and $m\neq 0$ components in adjacent Fs.  Hence the presence in Fig.~\ref{Figure2} of the {\it singlet} component in F" (solid blue line); despite its short-range nature in homogeneous systems, this component appears deep in the multilayer, at about $14\xi_c$ of the SF boundary.

The description above was made in terms of effective boundary conditions.  We point out that the features of Fig.~\ref{Figure2} could also be described in the quasi-classical diffusive limit in terms of the diffusion of pairs in the mixed state on either side of each interface.

The cascading effect and in particular the re-entrance of $m = 0$ components deep in the F multilayer has measurable consequences.  Considering the individual Gor'kov functions in the F" layer we note that the $m=0$ components are {\it largest} by an order of magnitude near the F'F" interface and are outweighed by the $m\neq 0$ as one moves towards the center of F". This change of the dominant character of the order parameter is seen in the Josephson current as one increases the thickness of the central F" layer.  The critical current flowing through the S$\nu$FS multilayer is calculated using \cite{champelPRL08,jcintegral}
\begin{eqnarray}\label{jcdensity}
j(x)&=&\frac{\pi T}{eR_N}\sum_{\omega_n\geq 0}\sum_{\alpha=0,y,z}\mathrm{Im}[f_{\alpha,n}\partial_zf^*_{\alpha,-n}],
\end{eqnarray}
where $R_N$ is the normal state resistance of the material and $e$ is the electron charge. In the configuration of Fig.~\ref{Figure1}, the Gor'kov vector $\mathbf{f}_n$ has no $x$ component.

\begin{figure}
\includegraphics[width=\columnwidth]{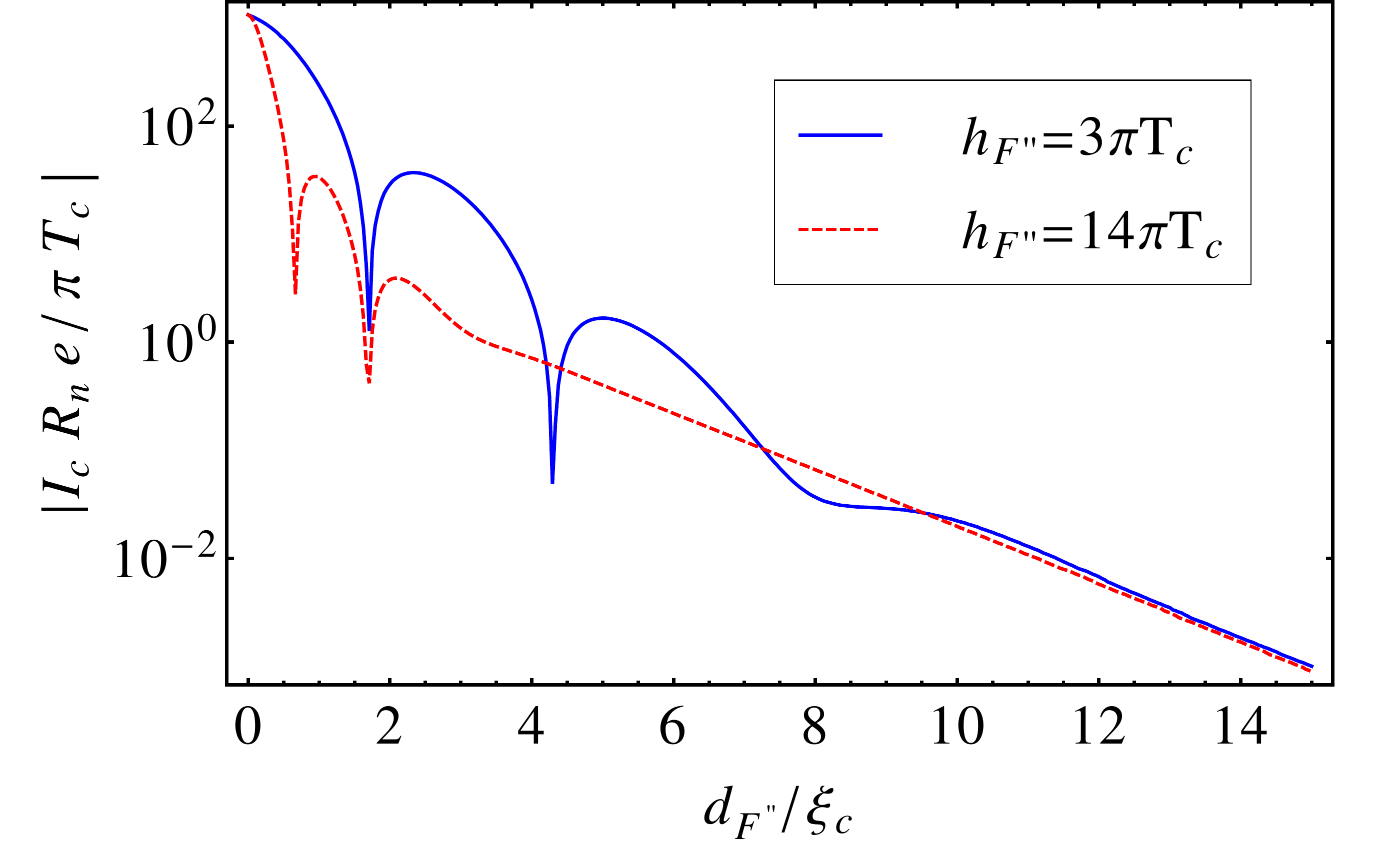}
\caption{\label{Figure4}
Josephson current through an S5FS junction as a function of the middle layer F" thickness for two magnetization strengths $h_{\rm F"}$. For small thicknesses ($d_{F"} \lesssim d_T$, see text), $m=0$ components dominate resulting in conventional $0-\pi$ transitions of the current. For large thicknesses ($d_{F"} \gtrsim d_T$), the $m\neq 0$ dominates and leads to a monotonous decay. Significant is that oscillations due to the {\it singlet} component in F" are found in such wide junctions. Same parameters as Fig.~\ref{Figure2}.}
\end{figure}

Figure \ref{Figure4} displays the current for two magnitudes of the magnetization in F" (Fig.~\ref{Figure1}). We identify two regimes and introduce a transition thickness $d_T$ to distinguish them; $d_T \approx 7\xi_c$ ($3\xi_c$) for the solid (dashed)  line. For  $d_{F"} \lesssim d_T$ the singlet component $f_0$ is the dominant contribution to the current and leads to conventional Buzdin-Bulaevskii $0-\pi$ transitions upon varying $d_{F"}$ \cite{buzdinJETP82}. By contrast, the $m\neq 0$ components dominate when $d_{F"} \gtrsim d_T$ and the expected monotonous decay is observed. It is surprising to find a $0-\pi$ transition for $d_{F"} \lesssim d_T$ given that we are dealing with a wide magnetic structure ($\sim 20\xi_c$).
 The very observation of this transition with increasing thickness of F" would constitute strong evidence for the presence of $m=0$ components deep in the F multilayer.

A series of conditions must be satisfied for cascade and Josephson current effects to be observed. The case of the S3FS system (spin valve) has been studied extensively both theoretically and experimentally \cite{houzetPRB07,khairePRL10,anwarPRB10}. The cascade effect is present in the trilayer as for any value of $\nu$. In such systems, however, only an $m\neq 0$ current is measured as indicated by the components of the order parameter in F' of Fig.~\ref{Figure2}; the $m\neq 0$ always dominates the $m = 0$ component. For this reason we considered a $\nu=5$ multilayer.

The findings exposed above lead to a few general remarks. The calculations were made for a multilayer with discontinuous rotation of the magnetization. The cascading effect will also occur in the more general case of a {\it continuously} rotating magnetization. The mixing then occurs {\it at each point} in the sample and the presence of both $m\neq 0$ {\it and} $m=0$ is thus found across the magnetic material, even in very wide magnetic films. Inhomogeneous magnetization nearly always produces a mixed state and the distinction between short- and long-range components looses its meaning.

The question is thus generally not whether one has a long-range triplet current in an inhomogeneous magnetic Josephson junction but, rather, how much of the current measured stems from the $m\neq 0$ component. Some systems, such as the exchange springs used in Ref.~\cite{bakerarXiv13} see a fraction of their currents stemming from the singlet ($m=0$) component despite being very wide.  The features discussed here may also not be specific to rotating magnetizations. It is conceivable that similar effects arise in normal metal multilayers with spin active boundaries and interfaces.

In conclusion, we analyzed the order parameter and Josephson current through a diffusive multilayer S5FS of misaligned homogeneous ferromagnets. The numerical solution of the Usadel equations reveals the existence of a cascade effect where $m=0$ and $m\neq 0$ transform between each other at the interfaces. We provided an exact solution of the non-linear Usadel equation in a homogeneous F to show that any state at its boundary necessarily evolves into a mixed state in F.  An experiment on a S5FS Josephson junction is proposed to reveal the existence of {\it singlet} components deep in the Fs. The main result of our work is that through the cascading effect {\it both} $m=0$ and $m\neq 0$ components are always present as a mixed state in a magnetic material with a rotating magnetization, irrespective of its width.

\section{Acknowledgments}

We gratefully acknowledge the support of the National Science Foundation (DMR-0907242, DMR-1309341) and the Army Research Laboratory. A.B.~thanks F.~Guinea and the Instituto de Ciencia de Materiales de Madrid (CSIC) for hospitality. T.E.B.~thanks the Graduate Research Fellowship at California State University Long Beach.



\begin{thebibliography}{99}
\bibitem{linksolAJP}
To include spin--flip times, we use the separatrix about the ejection angle, $\theta_e$, defined in Ref.~\cite{bakerAJP12}, and substitute the critical frequency $\omega_c^2 \to -\beta \in \mathbb{C}$ and the hoop's rotational frequency $\omega^2 \to 1/\tau_s$.
\bibitem{jcintegral}
The current density is nearly constant except within one coherence length of the SF interface in the wide limit. Hence, $I_c=\int_{-d_F/2}^{d_F/2} j(x)\,dx \approx j(0)\,d_F$, is the critical current per unit area at $\varphi=\pi/2$ \cite{zaikinZP81,vasenkoPRB08}.
\end{thebibliography}
\end{document}